# Enhancing human color vision by breaking binocular redundancy

Bradley S. Gundlach[1], Michel Frising[1,2], Alireza Shahsafi[1], Gregory Vershbow[3], Chenghao Wan[1,4], Jad Salman[1], Bas Rokers[5,6], Laurent Lessard[1], Mikhail A. Kats[1,4,6*]

[1]Department of Electrical and Computer Engineering, University of Wisconsin-Madison, Madison, WI
[2]Department of Mechanical and Process Engineering, ETH Zurich, Zurich, Switzerland
[3]Department of Art, University of Wisconsin-Madison, Madison, WI
[4]Department of Materials Science and Engineering, University of Wisconsin-Madison, Madison, WI
[5]Department of Psychology, University of Wisconsin-Madison, Madison, WI
[6]McPherson Eye Research Institute, University of Wisconsin-Madison, Madison, WI

**ABSTRACT**

**To see color, the human visual system combines the response of three types of cone cells in the retina—a compressive process that discards a significant amount of spectral information. Here, we present an approach to enhance human color vision by breaking its inherent binocular redundancy, providing different spectral content to each eye. We fabricated a set of optical filters that "splits" the response of the short-wavelength cone between the two eyes in individuals with typical trichromatic vision, simulating the presence of approximately four distinct cone types ("tetrachromacy"). Such an increase in the number of effective cone types can reduce the prevalence of metamers—pairs of distinct spectra that resolve to the same tristimulus values. This technique may result in an enhancement of spectral perception, with applications ranging from camouflage detection and anti-counterfeiting to new types of artwork and data visualization.**

**Introduction**

In the typical human eye, the three cone types—labeled "S" for short wavelengths, "M" for medium, and "L" for long—are sensitive primarily to light with wavelengths in the 390 - 530 nm, 400 - 670 nm, and 400 - 700 nm bands, respectively[1–4]. When excited by light, the signal from the cones is relayed though retinal ganglion cells, to the optic nerve, and then the brain, where it is further processed to produce a color sensation[5,6]. This process can be understood as a type of lossy compression from an $N$-dimensional spectrum, where $N$ is the number of wavelength bins necessary to sufficiently approximate a continuous spectrum, into a color, which is a three-dimensional object (Fig. 1). A manifestation of this $N$-to-three compression is metamerism, a phenomenon in which different spectra resolve to the same tristimulus values (*i.e.,* they appear as the same color, neglecting possible contextual effects)[2]. The number of cone types and the widths and separations of their spectral sensitivities govern the degree to which metamerism is a limitation of the visual system; for example, a hypothetical increase in the number of distinct cone types should result in a decrease in the prevalence of metamers. Cast in a signal processing perspective, an increase in the sampling rate of a spectrum (*i.e.*, the number of distinct cone types) improves the ability to detect sharp features in the spectrum[7].

Several studies have reported that a small percentage of humans, primarily women, express a mutated L cone in addition to the standard one, resulting in a total of four cone types, which may in principle enable vision with four color dimensions (tetrachromacy)[8–10]. Reports suggest that a few of these individuals can utilize this fourth photopigment type, and thus "perceive significantly more chromatic appearances" compared to typical, healthy humans with three cone types (trichromats)[11,12]. More broadly, it is reasonable to infer that an additional cone type would enhance spectral perception, provided subsequent neural processing can capitalize on its presence.

Here, we aim to simulate tetrachromatic (and possibly higher-dimensional) color vision in typical trichromatic humans by increasing the number of *effective* cone types in the visual system comprising the two eyes and a passive optical device. The term "effective" is used here to differentiate between true tetrachromatic vision, which would be defined by four distinct retinal photopigments that generate distinct neural responses. Our approach breaks the binocular redundancy of the two eyes, where the visual fields of each eye are overlapping, providing different spectral content to each eye via a wearable passive multispectral device comprising two optical transmission filters (Fig. 2).

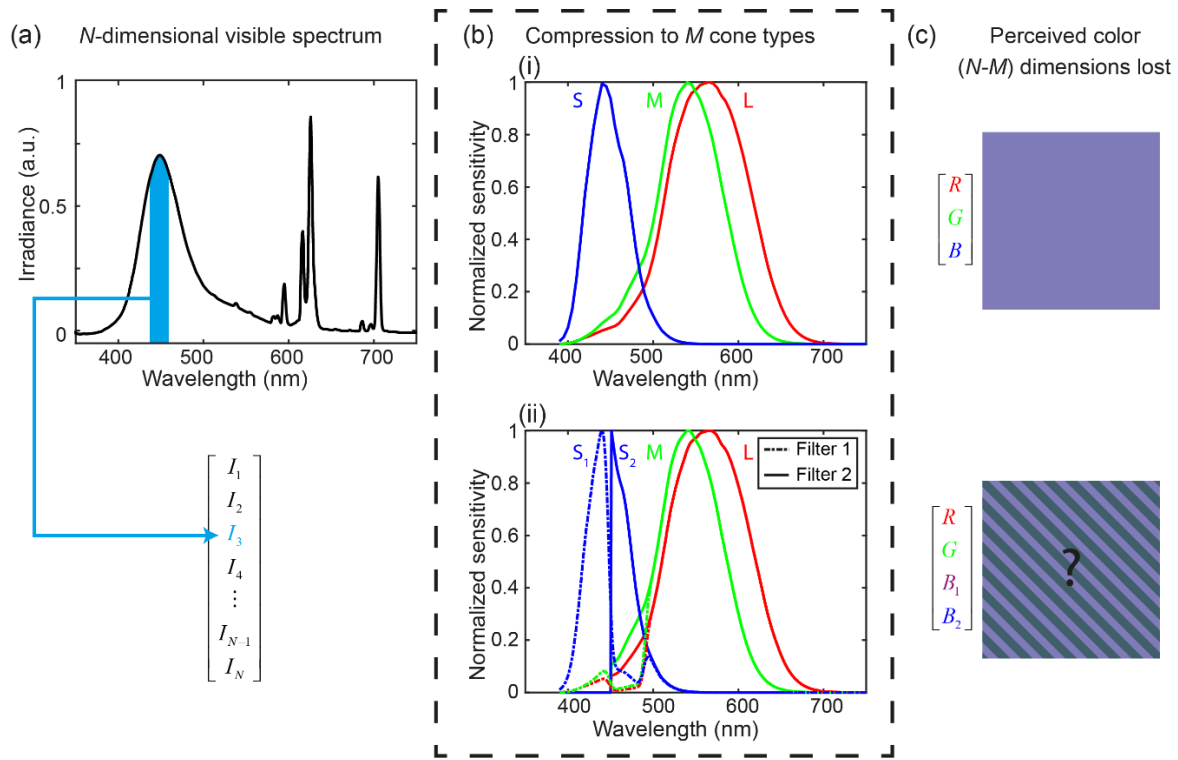

**Figure 1: Compression of spectral information (a)** A sample spectrum generated by a cathode ray tube (CRT) monitor displaying a purple color. The filled rectangle represents a single spectral bin, if the continuous spectrum is divided into $N$ bins. **(b)** (i) Normalized spectral sensitivity of the cone types for a typical trichromatic observer ($M = 3$). (ii) Normalized spectral sensitivity of the effective cone types for a typical trichromatic observer enhanced using our device ($M \cong 4$). **(c)** A representation of the perceived color of the spectrum in (a). The case with $M \cong 4$ cannot be displayed, but extra spectral information would be present compared to the $M = 3$ case.

A number of existing vision-assistive devices or techniques operate by breaking binocular redundancy, though usually in the spatial rather than spectral domain. Examples include hemianopia (partial blindness in the left or right visual field) treatment using spectacles with a monocular sector prism that selectively relocates the visual field in one eye, leaving the other eye unaffected, and thus conferring an additional 20° of visual-field sensitivity for binocular vision[13,14], and the treatment of presbyopia by correcting one eye for near vision and the other for distance vision[15].

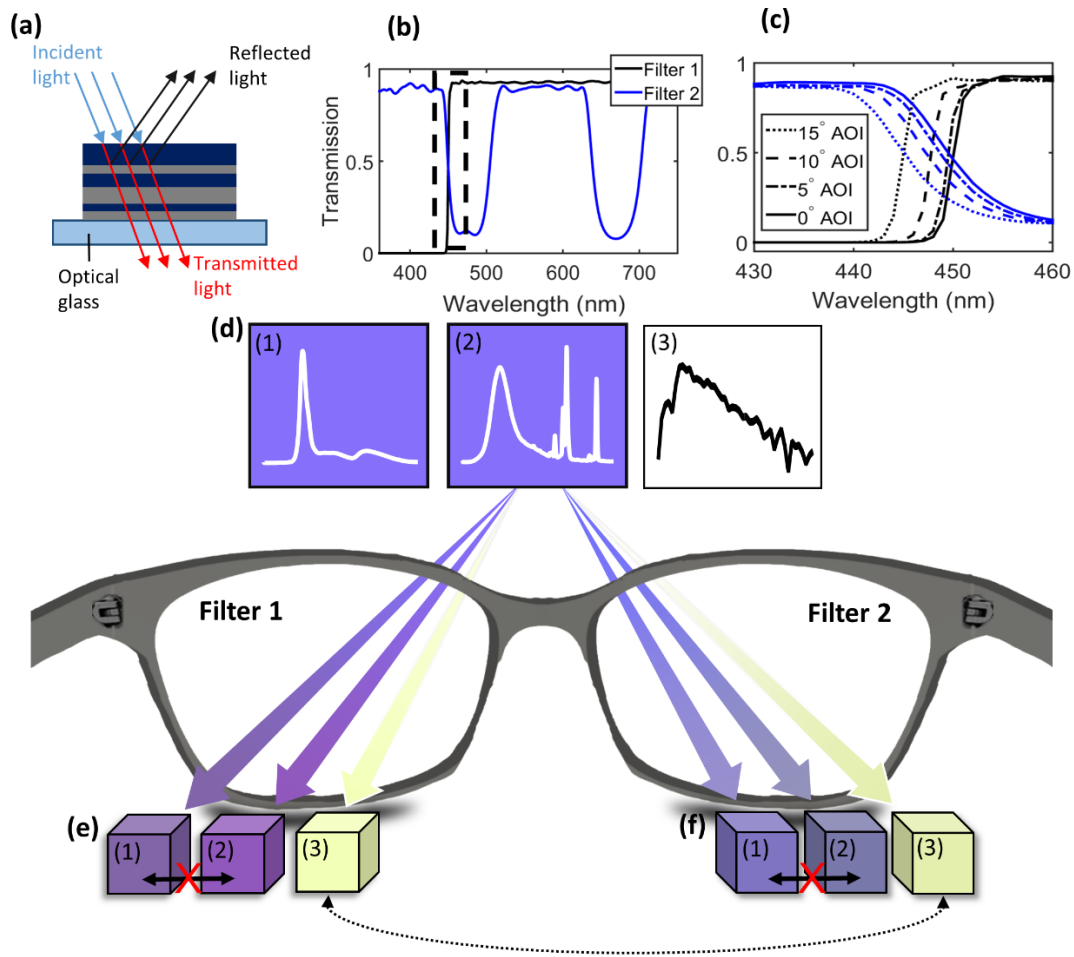

**Figure 2: Wearable passive multispectral device comprising two distinct transmission filters (a)** Simplified schematic of an optical filter comprising several thin-film layers. **(b)** Measured transmission spectra of fabricated Filters 1 (black) and 2 (blue). **(c)** Magnified portion of the transmission spectrum from (b), including angle dependence with angle of incidence (AOI) from 0 to 15°. **(d)** Colors of an example metameric pair (1, 2) and D65 broadband white light (3), as perceived by a typical trichromat. The traces in each box are the underlying unfiltered spectra (arbitrary units), taken from our experiments, as described in Fig. 3. **(e, f)** Rendered monocular colors of the spectra in (d) after passing through Filter 1 and Filter 2, respectively. Note that e(1) and e(2), f(1) and f(2) are substantially distinct, while e(3) and f(3) are similar in color due to the white-balance constraint enforced in our design.

We break binocular redundancy spectrally by using filters that selectively attenuate different wavelength bands to yield effective cone sensitivities (*i.e.,* the products of the cone sensitivities and the filter transmission spectra) that are different between the two eyes. This approach increases the number of effective cone types while preserving most spatial information. In this vein, the use of two simple band-pass filters was previously demonstrated to increase the dimensionality of color vision in dichromatic individuals (*i.e.,* those with two functioning cone types)[16]. Conversely, the goal in the present work is to enhance the dimensionality of a trichromat's visual system to beyond that of a typical human. Such an approach was briefly suggested by Cornsweet in 1970[17], but to our knowledge no specific design has been proposed or realized. We note that the use of even a single filter positioned in front of both eyes can help distinguish certain metamers[17], with the caveat that previously distinguishable spectra can become metamers when viewed through the filter; that is, a similar number of metamers (usually more) are created as are destroyed. In contrast, the use of two filters might be used to decrease the overall number of possible metamers. For this work, if at least one of the two filters can be used to differentiate a pair of spectra, we consider the pair to no longer be metamers.

**Results and Discussion**

*Filter Design and Construction*

The filter pair was designed using a standard psychophysical model to determine the perceived (monocular) colors corresponding to particular spectra[2,18]. The perceived colors were calculated using the International Commission on Illumination (CIE) 1931 2° standard-observer matching functions, and monocular color differences (*e.g.,* between colors 1 and 2) were calculated in the CIELAB color space using a standard color-difference metric (see *Methods* for further details)[2,19]:

$$\Delta E_{12} = \sqrt{(L_2 - L_1)^2 + (a_2 - a_1)^2 + (b_2 - b_1)^2} \qquad (\mathbf{1})$$

The filters were designed to enhance the ability of a typical trichromatic viewer to discriminate spectra while limiting adverse effects. For simplicity, we focused on a design that splits the response of the S cone, thus transforming the trichromatic visual system into one that simulates tetrachromatic vision. The S cone was chosen because its responsivity has relatively little overlap with those of the M and L cones (Fig. 1b(i)), so it can be attenuated while minimizing the impact on the effective responsivity (*i.e.,* the product of cone responsivity and the filter transmission response) of the other two cone types. To provide approximate parity between eyes, we partitioned the S cone responsivity such that each eye retains approximately half of the original response spectrum (Fig. 1b(ii)). Our secondary design goal was to ensure that the transmission of broadband white light (defined using CIE illuminant D65)[20] through the two filters results in similar tristimulus values. This constraint was put in place to minimize potential baseline disparities (*e.g.,* when viewing broadband white objects) between the eyes when the device is used in daylight. Though a particular implementation of this type of filter-based device generally depends on the illuminant chosen, the design presented here should work well for most illuminants along the Planckian locus[21].

The final device presented here comprises a 450 nm long-pass filter (Filter 1) and a 450 – 500 nm, 630 – 680 nm double-band-stop filter (Filter 2) (Fig. 2b-c); the filter designs were optimized by varying their stopband/passband positions and transmittances using simulated annealing to minimize the CIE ΔE color difference of D65 white light passing through Filters 1 and 2 (See *Methods* for further details). The 450 nm transition between Filters 1 and 2 is at the peak sensitivity of the S cone, and partitions it in half. However, due to the non-zero sensitivity of the M and L cones in the 450 – 500 nm region [Fig. 1b(i)], the M and L cones are also (unintentionally) attenuated by Filter 2. A second stopband at 630 – 680 nm was introduced to attenuate the effective responsivity of the M and L cones to broadband white light to preserve color balance. Though the filter designs were optimized for these constraints, we note that the design presented here is a proof of concept, and is not a unique or globally optimal solution.

To reduce cost and manufacturing time, an off-the-shelf component (450LP RapidEdge, Omega Optical), was used for Filter 1. The optimized transmission function of Filter 2 was realized using conventional thin-film technology[22], with alternating layers of silicon oxide ($SiO_2$, $n = 1.46$) and tantalum oxide ($Ta_2O_5$, $n = 2.15$) (Fig. 2a-b), deposited on an NBK7 glass substrate (see *Methods*). The two filters were then characterized by angle-dependent transmission spectroscopy, demonstrating that the transmission spectra are robust to incidence angles up to 5° away from the normal (Fig. 2c). Following fabrication, the filters were constructed into a pair of glasses.

*Experiments*

To test the performance of this design, we constructed a setup that generates metameric spectra using a liquid crystal display (LCD, True HD-IPS display on LG G3 smartphone) and a cathode ray tube (CRT, Dell E770P) monitor (Fig. 3a-b). The displays use different emission mechanisms, and thus produce distinct spectra when displaying the same color (See *Supplementary Information* for further analysis)[23,24]. Blocks of color generated by the displays were presented side by side using a 50/50 beam splitter, and the colors were individually adjusted until no perceivable color difference was present. The emission spectra of each monitor were recorded using a free-space spectrometer, allowing for chromaticity and color-difference calculations to be made given a standard observer. A threshold value of 2.3 for the CIE $\Delta E$ "just noticeable difference" was taken to define perceptually indistinguishable spectra (*i.e.*, $\Delta E < 2.3$)[25]. See *Methods* for further details of the experimental setup.

One representative example from this dual-display setup, using a pair of metamers that appear purple, is shown in Fig. 3b-d. Without the use of either filter, the two different spectra appeared as identical patches of color. However, when observed through either of the filters, the two can be differentiated.

Subjectively, we observed that, by looking at a particular patch through both filters simultaneously (*i.e.,* Filter 1 over the left eye, Filter 2 over the right), a color percept is observed that is different from the color perceived through either filter individually or with no filter. This new percept may be a manifestation of the approximately four effective cone types created by the pair of filters. We note that a related study involving dichromats demonstrated an increase in color dimensionality using band-pass filters, which the authors suggest the effect could be related to binocular lustre[16]. Our proposed capability to distinguish metamers by breaking binocular redundancy may be affected by binocular lustre and/or rivalry, which might be advantageous provided it can be used as a cue to a difference in hue. For example, a recent report has demonstrated that compression artifacts of pixels in virtual reality images can be easily detected due to lustre[26]. Note that lustre and rivalry are both dynamic phenomena, even for static stimuli[27–29]; thus the use of lustre/rivalry may result in a tradeoff between temporal and spectral resolution. Though our current experiments do not directly investigate the effect of lustre or rivalry, or probe to what degree the neuronal processing system of a trichromat can take advantage of the extra spectral information resulting from binocular filtering, this can be explored in future work.

We note that substantial differences in luminance between the two eyes, such as for spectra that transmit chiefly through only one filter, might lead to the Pulfrich effect[30]; however, our design was optimized to minimize differences in appearance of Illuminant D65, and the binocular luminance disparity required for the Pulfrich effect to occur is unlikely for most commonly occurring (*i.e.,* smooth/broad) spectra.

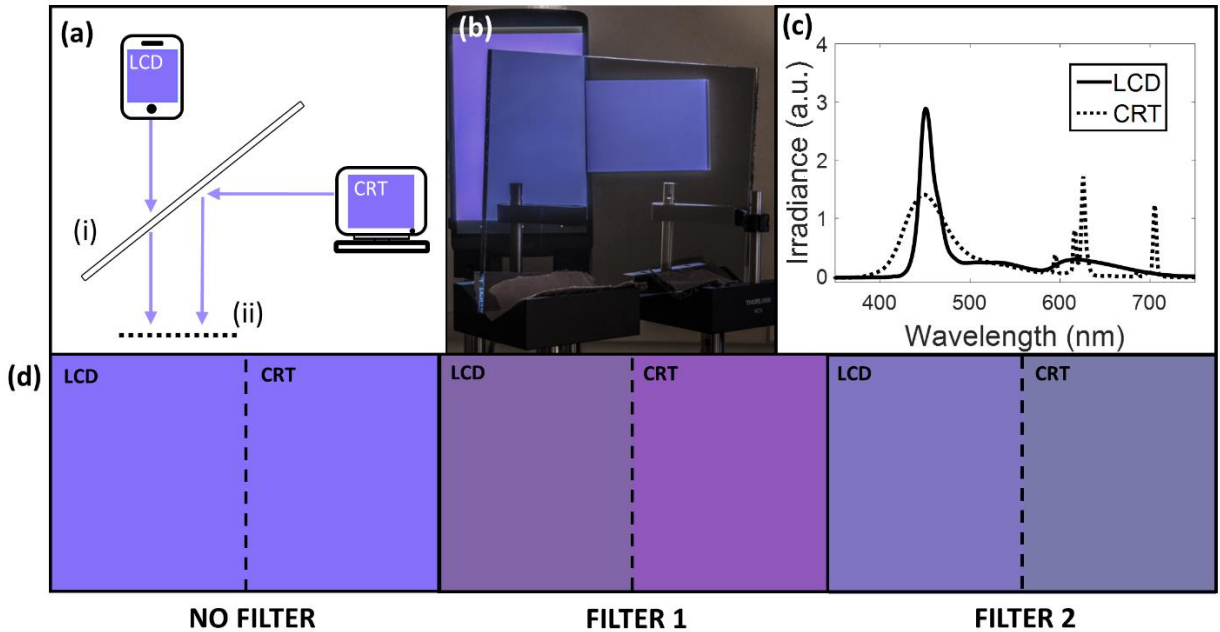

**Figure 3: Splitting metamers by breaking binocular redundancy (a)** Schematic of our metamer generation setup. Images from two monitors, an LCD and a CRT display, are combined using a 50/50 beam splitter (i), to be viewed at location (ii). The two monitors use different emission mechanisms, and thus generate different spectra for the same color. **(b)** Photograph taken at position (ii) in the schematic. **(c)** Measured emission spectra from the LCD (solid) and CRT (dashed) monitors while displaying the same purple color. Spectra are shown with arbitrary units. **(d)** Rendered colors of the spectra in (c) viewed through no filter, Filter 1, and Filter 2, respectively, showing that the metamers can be distinguished using either filter. See the supplementary information for discussion of the difference between rendered colors in (d) and those in the photograph in (b).

*Calculation of metamer reduction*

Broadly stated, the number of cone types and their frequency-dependent responsivities determines the extent to which metamerism is a limitation to the visual system. Our method increases the number of effective cone types, which should decrease the number of potential metamers, provided the subsequent neuronal processing can adapt appropriately (which seems to occur in the case of spatial multiplexing used for vision-assistive devices[13,14]). In general, quantitatively determining the decrease in the metamer frequency is difficult because the set of possible metamers is not bounded. Nevertheless, various metrics can be applied to roughly estimate this quantity. For this work, we developed two separate metrics that describe this decrease in metamer frequency given the following conditions. ***Condition 1***: Without the use of filters, a metamer pair is defined by two spectra with a color difference $\Delta E < 2.3$[25]. ***Condition 2***: With the use of binocular filters, such as those in Fig. 2b, a metamer pair is defined by two spectra with a monocular color difference $\Delta E < 2.3$ in each eye. That is, a pair of spectra is a metamer *if and only if* it is a metamer in each eye individually. We do not consider the possibility of other perceptual effects such as binocular rivalry or dichoptic color mixing.

Our first metric uses a Monte Carlo simulation to probe the effect binocular filters have on the perception of pairs of spectra, given the conditions above (Fig. 4). To start, a pair of reflectance spectra is generated by stochastically sampling intensity values from a uniform distribution at regularly spaced intervals within the visible wavelength range. Particularly, samples were taken uniformly at points: $\lambda_1, \lambda_2 ... \lambda_{N_s-1}, \lambda_{N_s}$, where $N_s$ is the total number of sampling points. The sharpness of the reflectance spectra was adjusted by changing $N_s$, with larger numbers leading to sharper features, and were interpolated at 10 nm intervals using a cubic spline to create smooth spectra. We assumed illuminant D65, and then filtered the reflected spectra by the filter transmission responses given in Fig. 2b. $\Delta E$ color differences were calculated between the pairs of spectra for the unfiltered case, and through Filter 1 and Filter 2. The method was performed for various number of iterations ($N_i$), which varied from 1,000 to 20,000,000, and the number of unfiltered ($M_u$) and filtered ($M_f$) metamers were recorded for each trial. We then defined a metric that represents the decrease in metamer frequency upon filtering:

$$P_m = \frac{M_u}{M_f}$$

For example, $P_m = 2$ represents a two-fold decrease in the number of metamers using the two filters. The results from this simulation, for several sampling values ($N_s$) and iteration numbers ($N_i$), are given in Fig. S5 of the *Supplementary Information*. Given the simulation conditions, the filters in this work result in up to a ~15× decrease in the number of metamers for randomly generated spectra; this effect appears to be greatest for moderately sharp spectral features ($N_s = 15$), and drops off for very broad or very sharp spectra. We note that, though the given metric seems to converge for larger iteration numbers (See *Methods* and *Supplementary Information*), these measures are only meaningful to within a factor of ~2 due to the stochastic nature of this calculation.

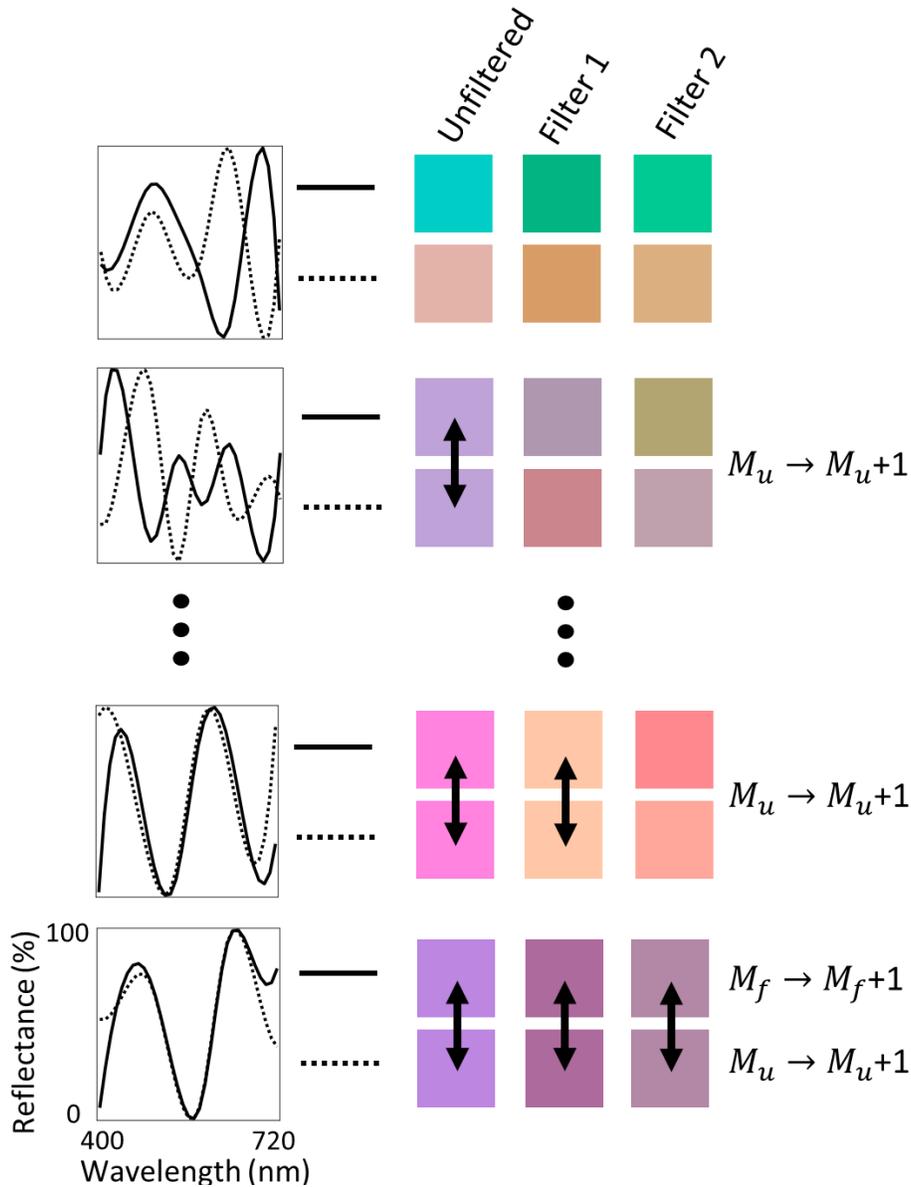

**Figure 4: Graphic representation of the Monte-Carlo metamer-reduction calculation**, where four pairs of randomly generated spectra are selected as an illustration. For the actual calculation, many pairs are generated. The corresponding rendered colors when viewed under illuminant D65 are shown for unfiltered, filter 1, and filter 2 cases, respectively. If a metamer is present in the unfiltered case, $M_u$ is incremented by 1. If metamers are present when viewed through *both* Filter 1 and Filter 2, $M_f$ is incremented by 1. Note that if a metamer exists when viewing through Filter 1 but not Filter 2, or vice versa, the spectra are considered distinguishable and $M_f$ is not incremented.

As further verification of the apparent decrease in metamer frequency, we also developed a more-abstract mathematical method (See *Supplementary Information* for a complete description of this calculation, abridged here for clarity). Rather than comparing stochastically generated spectra, as above, this method aims to calculate the *overall* number of spectra that map to perceptually indistinguishable tristimulus values ($\Delta E < 2.3$). For a given reference point in LAB space, $[L_o, a_o, b_o]$, the number of metamers (with respect to the reference point) was determined by counting the spectra, $I(\lambda)$, that map to LAB coordinates within a sphere of radius 2.3 around the reference point. We determined the number of metamers by calculating the volume of spectra, represented by an ellipsoid in $N$-dimensional space, where $N$ is the number of discrete wavelength bins that define a spectrum. However, calculating the exact volume of high dimensional ellipsoids in this case is difficult; instead, we calculate the volume of the max-inscribed ellipsoid subject to box constraints, which represents an upper-bound of the true value and is more computationally efficient[31] (see *Supplementary Information* for more details). The volume of this ellipsoid represents the number of metamers, for a given reference point, for the unfiltered case ($V_u$). For the filtered case, the union of two ellipsoids, corresponding to each filter individually, represents the number of metamers ($V_f$); this is equivalent to **Condition 2** above, where we assume that monocular metamerism must be present in both eyes simultaneously to yield indistinguishable color percepts in the filtered case. Thus, the overall decrease in metamer frequency is given by:

$$F_m = \frac{V_u}{V_f}$$

Where $F_m = 2$, as an example, represents a two-fold decrease in metamer frequency. This process was repeated for 500 LAB reference points from randomly generated spectra to adequately sample the color space. The number of wavelength samples ($N_s$) was also varied to again explore the effect of spectral sharpness; as in the Monte-Carlo simulation, the decrease in metamer frequency occurs around 12 – 16 bins. Using this metric, we estimate a decrease in metamer frequency by one-to-two orders of magnitude when using our passive multispectral device (see Table S1 in the *Supplementary Information*). By the same metric, a single-filter system designed to improve vision in color-vision-deficient individuals[32] seems to provide no decrease in the frequency of apparent metamers.

**Conclusion**

By breaking the inherent chromatic redundancy in binocular vision, our method provides the user with more spectral information than he/she would otherwise receive. In the present design, the S cone is partitioned using a pair of filters that results in photoreceptor responses consistent with a visual system that utilizes approximately four cone types (i.e., simulated tetrachromacy). The current demonstration splits the S cone, which is more selectively sensitive to blue-colored objects, and might find direct applications such as differentiating structural color versus natural pigments (See *Supplementary Information*)[33]. However, it is possible to use similar methods to design filters that more strongly affect metamers that appears as green and red, that are more prevalent in nature[34]. While the possibility of natural tetrachromacy in a fraction of the population has received both academic and popular interest[10–12], the technology demonstrated here has the potential to simulate tetrachromatic vision in anyone with typical, healthy trichromatic vision. The extent to which observers can (or can learn) to take advantage of the additional spectral information is yet to be determined.

Given two eyes and three types of cones, it should be possible to increase the number of effective cones up to six using our approach, and potentially even more with spatial or temporal multiplexing. It may also be possible to generate personalized designs to improve color discrimination for individuals with color-vision deficiencies. This technology can be integrated in a simple pair of eyeglasses or sunglasses, and could have immediate applications in camouflage detection, quality control, anti-counterfeiting, and more. More broadly, the ability to see many more colors has intriguing opportunities for design and artwork, and for data representation with extra color channels.

**Methods**

<u>Color calculations and CIE color differences.</u> The International Commission on Illumination (CIE) standard was used for color calculations, represented by the equation[2,18]:

$$\Theta = \int_{\lambda_1}^{\lambda_2} \bar{\theta}(\lambda)T(\lambda)I(\lambda)d\lambda \Big/ \int_{\lambda_1}^{\lambda_2} \bar{y}(\lambda)I(\lambda)d\lambda$$

where $\Theta = [X; Y; Z]$ are the XYZ tristimulus values, $\bar{\theta}(\lambda) = [\bar{x}(\lambda), \bar{y}(\lambda), \bar{z}(\lambda)]$ are the 1931 CIE 2° standard observer matching functions, $T(\lambda)$ is the transmission spectrum of the filter, and $I(\lambda)$ is the spectral irradiance of light passing through the filter. The XYZ tristimulus values can be transformed to a different color space (e.g., RGB); here, we use the CIELAB color space because it is more perceptually uniform and allows for straightforward calculations of perceived color differences. The XYZ to LAB transformation is given by[19]:

$$L_1 = 116 f\left(\frac{Y}{Y_n}\right) - 16, a_1 = 500\left[f\left(\frac{X}{X_n}\right) - f\left(\frac{Y}{Y_n}\right)\right], b_1 = 200\left[f\left(\frac{Y}{Y_n}\right) - f\left(\frac{Z}{Z_n}\right)\right],$$

where

$$f(t) = \begin{cases} t^{1/3} & , \quad t > \left(\frac{6}{29}\right)^3 \\ \frac{1}{3}\left(\frac{6}{29}\right)^2 t + \frac{4}{29}, & otherwise \end{cases}$$

and $X_n, Y_n, Z_n$ are the tristimulus values of the reference white point. Here, white light is defined by the CIE D65 standard illuminant, which roughly corresponds to average mid-day solar illuminance. The white point of D65 is (95.047, 100.000, 108.883)[20].

<u>Filter Design</u>

An iterative optimization approach was used to design the transmission of Filter 1 and Filter 2, where the filter responses became more complex as our intuition grew between iterations. This approach was used to meet the primary design goal, splitting the spectral response of the S cone between eyes, while also enforcing other optimization conditions such as a perceptual color balance for D65 white light between eyes.

The final filter designs comprise a 450nm longpass filter (Filter 1) and a 450-500nm, 630-680nm double bandstop filter (Filter 2). The 450nm transition region, where Filter 1 cuts on and the first bandstop of Filter 2 cuts off, occurs roughly at the peak sensitivity of the S cone. Therefore, Filter 1 transmits the long wavelength half of the S cone, while the first stopband of Filter 2 transmits the short wavelength half of the S cone. However, due to the nonzero sensitivity of the M and L cones between 450-500nm, their sensitivities are also inadvertently attenuated, impacting the D65 color balance between eyes. The second bandstop of Filter 2, between 630-680nm, attenuates the long wavelength tails of the M and L cone sensitivities, restoring color balance between eyes. A 450nm longpass filter (Omega Optical, 450LP RapidEdge) was chosen as Filter 1. Filter 2 was optimized using constrained optimization by linear approximation (COBYLA) to minimize the merit function: $\Delta E/Width_{bandstop}$, where $\Delta E$ is the color difference between Filter 1 and Filter 2 when transmitting D65 white light and $Width_{bandstop}$ is the spectral width of the short-wavelength band-stop region of Filter 2. This merit function ensures satisfactory D65 color balance between eyes while also maximizing the difference between filters, enhancing their ability to distinguish spectra. For Filter 2, the transmittance of the pass and stop-bands were constrained between 5-15% and 80-95%, respectively, and the longer wavelength stopband was constrained between 600-700nm to prevent attenuation of the M and L cones at their peak sensitivities (~550nm, ~580nm respectively). This procedure yielded an optimized response for Filter 2 with stopbands at 450 - 500 nm and 630 - 680 nm, and stopband/passband transmittance of 10% and 90%, respectively; the color difference for illuminant D65 between Filter 1 and Filter 2 is $\Delta E = 5.21$, with chromaticities of (0.348,0.406) and (0.35,0.415), respectively.

### Thin-film optimization and construction

The required film thicknesses were determined by conventional thin-film optimization methods, including gradual evolution[35] and needle optimization[36], to implement the target transmission function. The final stack was constrained to be less than 75 total layers, and each layer between 10 and 500 nm thick. The filter was optimized such that the transmission would not change significantly for incident angles up to 5° away from the normal. The films were deposited using ion-assisted sputtering onto an NBK7 glass substrate at a thin-film foundry (Iridian Spectral Technologies, Ontario, Canada). See *Supplementary Information* for more information about the thin-film design.

### Metamer Generation

An LCD (True HD-IPS on LG G3 smartphone) and CRT (Dell E770P) display were used to generate metameric pairs. The monitors were placed at a 90° angle from one another, and a large 50/50 beam splitter (Edmund Optics) was placed at 45° between the displays such that images from the two displays could be projected directly next to each other with no border. To find a metamer, a block of color was displayed on the CRT display, and a user-controlled 3-axis joystick was used to adjust the LCD image until no perceivable color difference was detected by the observer; the 3-axis joystick controlled colors in the HSV color space. The entire experimental setup was enclosed in a wooden box, painted black on the inside, and square apertures were placed on each display to ensure the images were displayed with black backgrounds to mitigate possible contextual perception effects[28]. Spectra from each monitor were acquired using a free-space spectrometer (Ocean Optics FLAME VIS-NIR with cosine corrector), normal to and adjacent to each display screen. Spectra for the white point of each monitor are shown in the S*upplementary Information*.

### Monte Carlo Simulation

Reflectance spectra were calculated by generating random values at a defined number of sampling numbers ($N_s$) within the visible wavelengths (400 – 720 nm) using Matlab's rand function; $N_s$ was varied between 4-35 points to define the sharpness of spectral features, and the spectra were interpolated using a cubic spline. CIE 1931 2° matching functions were used to calculate tristumlus values for illuminant D65 reflected from the objects. Illuminant D65 was used as the white-point for conversion to the CIELAB space. A threshold of $\Delta E < 2.3$ was used to define indistinguishable tristimulus values[25]. The simulation was performed for several number of iterations ($N_i$), from 1,000-20,000,000, to determine if the defined metric $P_m$ converged for a given $N_s$. For $N_i$ greater than 1,000,000 values for $P_m$ converged to values within ~20% of each other within a given $N_s$.

### More-abstract calculation of metamer frequency

The volume approximation ratios were computed using CVX, a package for specifying and solving convex programs[37]. Reference points $[L_0, a_0, b_0]$ were chosen by uniformly sampling discretized spectra and mapping them to LAB tristimulus values. The computation was performed for various wavelength binning values ($N_b$), between 7-18, to vary the broadness/sharpness of spectra, and 500 reference points were computed for each binning value. A detailed summary of this calculation, including a formalized mathematical treatment, can be found in the *Supplementary information.*

**Data Availability Statement** All data generated or analyzed during this study are included in this published article (and its Supplementary Information files).

**Acknowledgments**. B.G. is supported by the National Science Foundation Graduate Research Fellowship under Grant No. DGE-1256259. This work was partially supported by startup funds from UW Madison, and partially by the AFOSR (FA9550-18-1-0146). We also thank Middleton Spectral Vision for access to their hyperspectral imaging systems.

**Competing Interests.** UW-Madison has filed a patent based on the technology presented in this manuscript, Pub. No: US2018/0041737 A1. B.G. and M.K. are listed as co-inventors on this patent.

**Author Contributions.** M.K. and B.G. developed the concept. B.G. performed the design and optimization, and carried out the majority of the experiments. B.R. assisted with brainstorming and experimental design. M.F., A.S., C.W., J.S., and G.V. assisted with the experiments. B.R., M.F., and L.L. assisted with analysis and interpretation. L.L. performed the "more abstract" metamer reduction calculation. B.G. and M.K. wrote the manuscript, with input from all authors. M.K. supervised the research.

# Supplementary Information

# Enhancing human color vision by breaking binocular redundancy

Bradley S. Gundlach[1], Michel Frising[1,2], Alireza Shahsafi[1], Gregory Vershbow[3], Chenghao Wan[1,4], Jad Salman[1], Bas Rokers[5,6], Laurent Lessard[1], Mikhail A. Kats[1,4,6*]

[1]Department of Electrical and Computer Engineering, University of Wisconsin-Madison, Madison, WI
[2]Department of Mechanical and Process Engineering, ETH Zurich, Zurich, Switzerland
[3]Department of Art, University of Wisconsin-Madison, Madison, WI
[4]Department of Materials Science and Engineering, University of Wisconsin-Madison, Madison, WI
[5]Department of Psychology, University of Wisconsin-Madison, Madison, WI
[6]McPherson Eye Research Institute, University of Wisconsin-Madison, Madison, WI

Narrative explanation of the filter design process

The filters in this work were designed using a white-balance condition that ensures that broadband "white light" that passes through the two filters is perceived similarly, to prevent significant color clashing between the two eyes under typical viewing conditions. Simultaneously, the filters are designed to be sufficiently distinct, which results in each eye receiving different spectral information. We used a design approach where each revision increased in complexity, building intuition at each design stage.

For the first revision, a brick-wall longpass filter and a band-stop filter were used to split the blue cone response without significantly affecting the other cone types (Fig. S1). The filters were constrained such that the longpass filter cut-on wavelength was equal to the band-stop filter cut-off wavelength, which ensured that at least one eye was sensitive to every region of the visible wavelengths (*i.e.,* no wavelength was attenuated by both filters). The band-stop filter cut-on wavelength was chosen to be 450 nm to minimize the effect of the filters on the M and L cone responses; thus, we were left to decide the longpass cut-on and band-stop cut-off wavelengths (which are enforced to be equal), and which must be below 450nm. With these constraints, the position of the cut-on/cut-off wavelength was optimized in order to minimize the CIE $\Delta E$ color difference between D65 white light passing through each filter. Figure S1(a) shows the result of this optimization, with a minimum $\Delta E$ color difference of 0.596 with the cut-on/cut-off wavelength at 437 nm. A $\Delta E$ color difference less than 2.3 typically means that the colors are indistinguishable. The resulting optimized filters are: a longpass filter (Filter 1) with a 437 nm transition wavelength, and a band-stop filter (Filter 2) with a stopband of 437 - 450 nm. Figure S1(c) shows the rendered color of D65 white light through each filter, and demonstrates the excellent white-balance between filters, because the color samples are indistinguishable. However, the band-stop width of the resulting Filter 2 is only 13 nm, quite low in comparison to the ~300 nm range of the visible wavelengths. In order to effectively differentiate spectral features in everyday scenes, which are typically quite broad (>10 - 20 nm), the band-stop width must be larger.

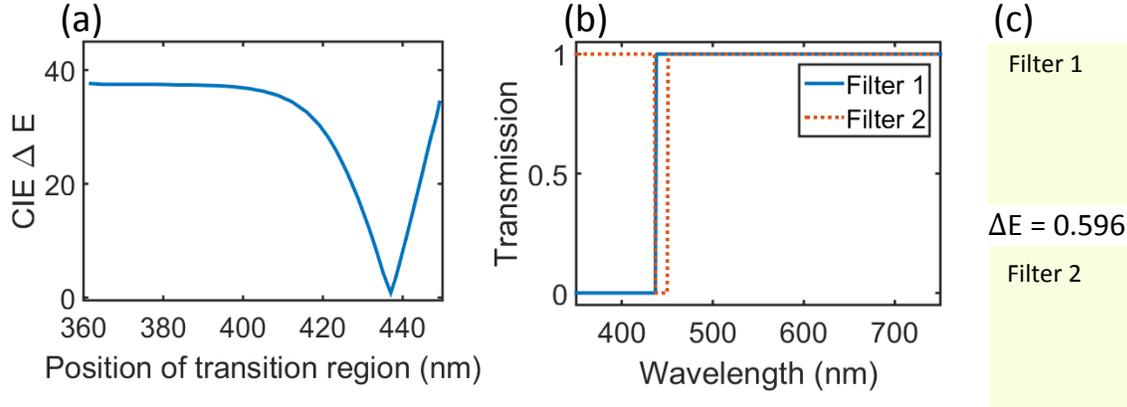

**Figure S1:** (a) Plot showing the CIE ΔE color difference between filters 1 and 2 when transmitting CIE D65 white light versus cut-on/cut-off wavelength. (b) Optimized transmission response for filters 1 and 2 after the first design revision, with a band-stop region between 437 – 450 nm. (c) Rendered colors for CIE D65 white light transmitted through each filter from (b), with an optimized ΔE color difference of 0.596.

Because this initial design produced a filter set with good white-balance, the same general approach was used in the following design revision. A similar optimization was performed as above, but the bandstop width of Filter 2 was constrained to be larger than 25 nm; this width was chosen such that it was larger than many typical spectral features found in nature, which could therefore be resolved with these filters. Instead of minimizing just the $\Delta E$ color difference between the filters, a modified merit function (MF) was used that also accounted for the filter width: $MF = \Delta E/Width_{bandstop}$. A brute-force optimization was performed, which varied the cut-on and cut-off wavelengths of the two filters (again with the longpass cut-on and bandstop cut-off wavelengths being equal). The transmission spectra of the optimized filters are given in Figure S2: a longpass filter (Filter 1) with a 450 nm cut-on wavelength, and a bandstop filter (Filter 2) with a stopband of 450 – 500 nm. The resulting filter set has a $\Delta E$ color difference of 14.15, significantly above the 2.3 just noticeable difference threshold. Therefore, it is clear that for the simple design using a longpass and single-bandstop filter, there is a tradeoff between white-balance and width of the band stop region.

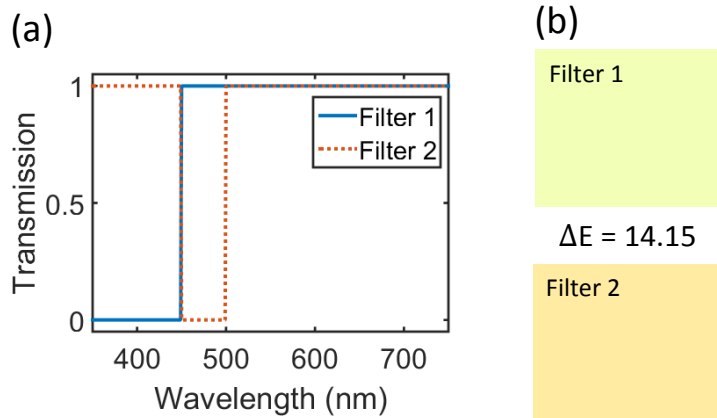

**Figure S2:** (a) Optimized transmission response for filters 1 and 2 after the second design revision, with a band-stop region of 450 - 500 nm. (b) Rendered colors for CIE D65 white light transmitted through each filter from (b), with an optimized ΔE color difference of 14.15

By considering the cone responses (Fig. 1c), it is clear why this filter design results in worse white balance. Beyond 450 nm, the longpass filter allows all wavelengths of visible light through; the M and L cones have very little response below 450 nm. However, the band-stop filter attenuates wavelengths between 450 and 500 nm, where the M and L cones have significant sensitivity. Therefore, the two filters affect the M and L cones very differently, which results in poor color balance. In particular, the band-stop filter (Filter 2) introduces a slight red tint compared to the longpass filter (Filter 1), which is also clear from the rendered colors in figure S2. Therefore, in order to maintain the high band-stop width of Filter 2 while improving the white balance of the filter pair, a second stopband was introduced to Filter 2 in order to soften the "red response" of the previous design. In this design, the shorter wavelength band-stop region of Filter 2 splits the S cone, whereas the longer wavelength band-stop region improves the white balance of the filter pair.

In the final design revision that was implemented and described in the main text, the response of the longpass filter (Filter 1) was not optimized any further; in fact, the transmission response of a commercially available 450 nm longpass filter was used to define Filter 1. This was done to decrease manufacturing costs, such that only Filter 2 required custom design and manufacturing. The short-wavelength band-stop region of the Filter 2 was also used from before (allowed to change only slightly during optimization), and a preliminary long wavelength band-stop region was added between 600 and 700 nm, to be optimized further. An error function was also implemented to smooth the transition regions; in the previous design revisions, the transition regions had a sharp vertical slope, which is difficult to achieve in practice. The smoothness of the filters can be adjusted by changing the proportionality constant (a) of the error function: $y = \mathrm{erf}(ax)$.

Unlike the previous revisions, Filter 2 was optimized using a more rigorous method compared to the brute force method used above. A constrained optimization by linear approximation (COBYLA) method was implemented in a stochastic basin-hopping algorithm to optimize the position of the cut-on/cut-off wavelengths, transmission of the pass and stop-bands, and slope of the transition regions. The transition regions of the filter's short-wavelength band-stop region was constrained within +/- 10 nm of their previous values; this was done to maintain the overall shape of the previous design while allowing some room for color balance optimization. The transmittance was constrained between 5 and 15% in the stopbands, and between 80 and 95% in the passbands, to allow for high throughput and relative ease of manufacturing. The error function proportionality constant was constrained between 0.25 and 1. The long-wavelength stopband region was constrained between 600 – 700 nm, but the band-stop width was not constrained; this was done to prevent attenuation of the M and L cones at their peak sensitivities (~550nm, ~580 nm respectively), while also preventing needless optimization beyond the visible wavelengths (> 700 nm). With these constraints in place, Filter 2 was optimized in order to minimize the modified merit function: $\Delta E/Width_{bandstop}$, where $\Delta E$ is the color difference between Filter 1 and Filter 2 when transmitting D65 white light and $Width_{bandstop}$ is the spectral width of the short-wavelength band-stop region of Filter 2. This procedure yielded an optimized response for Filter 2 with stopbands at 450 - 500 nm and 630 - 680 nm, and stopband/passband transmittance of 10% and 90%, respectively (Figure S3(a)). The rendered color of transmitted D65 white light through the filters is given in Figure S3(b), with a $\Delta E$ color difference of 5.21.

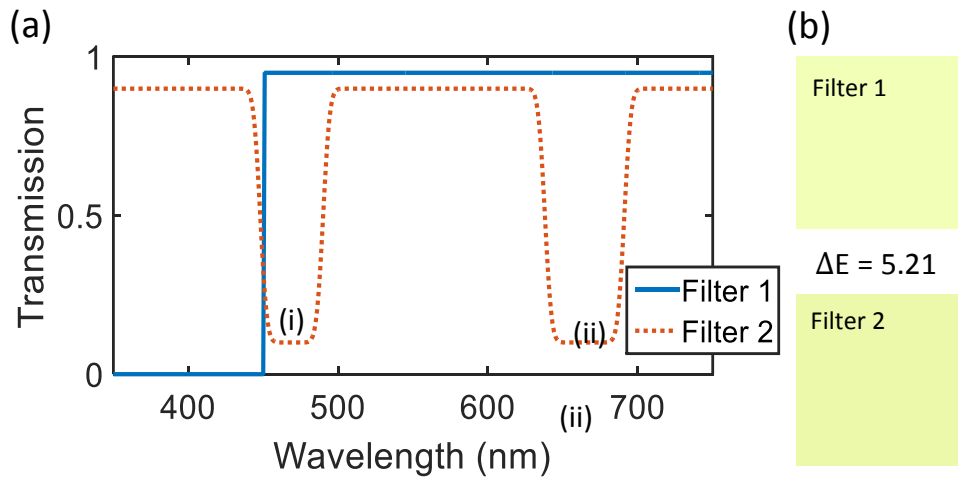

**Figure S3:** (a) Optimized transmission response for filters 1 and 2 after the third design revision, with band-stop regions between 450 – 500 nm (i) and 630 – 680 nm (ii). Region (i) splits the short-wavelength cone, region (ii) improves the white-balance of the filter pair (b) Rendered colors for CIE D65 white light transmitted through each filter from (a), with a ΔE color difference of 5.21.

Methodology for counting metamers in "more abstract" calculation:

In this section, we describe the technique by which we estimated the reduction of the occurrence of metamerism using our vision enhancement device.

Each spectrum $I(\lambda)$ is mapped to LAB tristimulus values $[L, a, b]$ via the CIE matching functions described earlier in the supplementary. Counting the number of metamers for a particular LAB reference point $[L_0, a_0, b_0]$ amounts to counting the number of different spectra $I(\lambda)$ that map to tristimulus values that are within a sphere in LAB space of radius $\Delta E$ of the reference point. The number of metameric spectra is infinite, so we instead compute a surrogate quantity. Roughly, we discretize each spectrum by wavelength so each spectrum can be abstracted as a point in a finite-dimensional space. We then count metamers by computing the volume that they occupy in this space. The details of the computation are described below.

1. Represent spectra by using $N_S$ equally spaced samples in wavelength. For example, $I(\lambda)$ is represented as a vector $[I_1, I_2, \ldots, I_{N_S}]$, which corresponds to the intensities at the wavelengths $[\lambda_1, \lambda_2, \ldots, \lambda_{N_S}]$.
2. The map $[I_1, \ldots, I_{N_S}] \to [L, a, b]$ from the discretized spectrum to LAB tristimulus values is smooth and nonlinear. Since the map is only being evaluated in a local neighborhood of the reference point, the map is well approximated by its first order Taylor expansion. This allows us to replace the nonlinear map with an affine function $g(I_1, \ldots, I_{N_S}) = [L, a, b]$.
3. Let $S_0 = \{ [L, a, b] \mid (L - L_0)^2 + (a - a_0)^2 + (b - b_0)^2 \leq \Delta E^2 \}$ be the set of tristimulus values indistinguishable from the reference point. The set of metameric spectra is given by the image of $S_0$ under the inverse map $g^{-1}$.
4. Since the inverse map $g^{-1}$ is affine and the set $S_0$ is a sphere in LAB space, the image $g^{-1}(S_0)$ is an ellipsoid in the discretized spectrum space [S1]. Note that this ellipsoid is degenerate; it will be infinite in the directions corresponding to the kernel of $g$.
5. We assume that we are counting "reflection metamers" or "transmission metamers" under a certain illuminant. That is, we are excluding metamers generated by active emissive sources for the sake of this calculation, since including unbounded emissive sources complicates this calculation further. Under this assumption, the allowed intensities are not infinite, since every admissible spectrum has intensities bounded by the corresponding intensities of the illuminant. More formally, define the set of admissible spectra as $C_0 = \{ [I_1, \ldots, I_{N_S}] \mid 0 \leq I_k \leq I_k^{D65} \text{ for } k = 1, \ldots, N_S \}$, where $I_k^{D65}$ is the intensity at $\lambda_k$ of the D65 illuminant.
6. The volume of metameric spectra is therefore the volume of the set $g^{-1}(S_0) \cap C_0$.

If the spectrum is filtered through a filter $T(\lambda)$, the map $g$ must be replaced by a map $g_T$ that accounts for the filter $T$. The derivation is otherwise identical; metameric spectra are given by the set $g_T^{-1}(S_0) \cap C_0$. If the spectrum is filtered through $T_1(\lambda)$ for one eye and $T_2(\lambda)$ for the other eye, spectra are only counted if they are metameric for *both* eyes. This results in the set $g_{T_1}^{-1}(S_0) \cap g_{T_2}^{-1}(S_0) \cap C_0$.

We can compare configurations (e.g. natural human vision versus vision augmented by our device, or vision augmented by two different filters sets) by comparing the volumes of their respective metameric spectra. For example, to compare the unfiltered case (natural human vision) to the case of vision modified by our two-filter passive multispectral device, we would compute the ratio:

$$\rho = \frac{\text{Vol}(g^{-1}(S_0) \cap C_0)}{\text{Vol}(g_{T_1}^{-1}(S_0) \cap g_{T_2}^{-1}(S_0) \cap C_0)}$$

A ratio of $\rho = 20$ would signify that metameric spectra are 20 times less abundant when the two-filter passive multispectral device is used as compared to the unfiltered case. Specifically, if spectra are sampled from a uniform distribution on intensities, a spectrum is 20 times less likely to be metameric.

Computing the ratio $\rho$ is challenging because the volumes involved have irregular shapes; they are intersections of degenerate (high-dimensional) ellipsoids with box constraints. In order to approximate the ratio $\rho$, we approximate each volume by the volume of its max-volume inscribed ellipsoid. An illustration of a max-volume inscribed ellipsoid is shown below (Fig. S4).

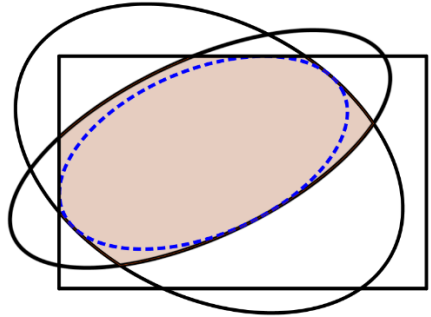

**Figure S4:** The shaded region is the intersection of ellipsoids and box constraints. The area is approximated by the max-area inscribed ellipsoid (dotted curve).

It turns out the max-volume inscribed ellipsoid can be efficiently computed using semidefinite programming techniques. See for example [S2]. We approximated the ratio $\rho$ by using the inner ellipsoid approximation for both the numerator and denominator. Volume approximation ratios were computed using CVX, a package for specifying and solving convex programs [S3]. We computed approximate volume ratios for the two-filter case as well as the one-filter case, using the transmission spectrum of a filter sold by EnChroma for alleviating some of the adverse effects of red-green color vision deficiency (EnChroma filter [S4]). In each case, we tried several different discretization points $N_S$ and we repeated the computation for 500 different reference points $[L_0, a_0, b_0]$. The reference points were selected by choosing discretized spectra at random, with reflectance values sampled from a uniform distribution between 0 and 1 (*i.e.*, $N_S$ randomly sampled values per spectrum), and mapping them to LAB tristimulus values. A summary of the results is shown in Table S1. Using two filters results in a dramatic decrease in metamers, roughly consistent over the range of tested discretizations. In contrast, using a single filter (EnChroma in this case) has little effect on the number of metamers.

| Discretization points ($N_S$) | 7 | 9 | 12 | 14 | 16 | 18 |
|---|---|---|---|---|---|---|
| Two filters, mean | 22.977 | 115.823 | 120.580 | 160.634 | 178.020 | 111.243 |
| Two filters, median | 17.251 | 80.690 | 57.4080 | 85.885 | 77.153 | 41.369 |
| EnChroma, mean | 0.945 | 0.887 | 1.090 | 0.944 | 1.125 | 1.086 |
| EnChroma, median | 0.945 | 0.870 | 1.082 | 0.944 | 0.924 | 1.017 |

**Table S1 –** Results of the metamer ratio approximation. Mean and median $\rho$ values are computed over 500 randomly generated spectra for each discretization. Using two filters reduces the frequency of metamers by a factor of about 50 on average, while using a single EnChroma filter has a negligible effect on the frequency of metamers

Monte Carlo Metamer Calculation

The Monte Carlo simulation, as described in the main text, was performed for several values of spectral sharpness $N_s$, with higher numbers signifying sharper features, and number of iterations $N_i$ (Fig. S5). As discussed in the main text, $P_m$ is largest for moderately sharp spectral features ($N_s = 15$). The metamer reduction metric seems to converge to within ~20% of neighboring values when $N_i$ reaches ~1,000,000. Note there are missing values for low number of iterations because the sample size was not large enough to generate metameric spectra.

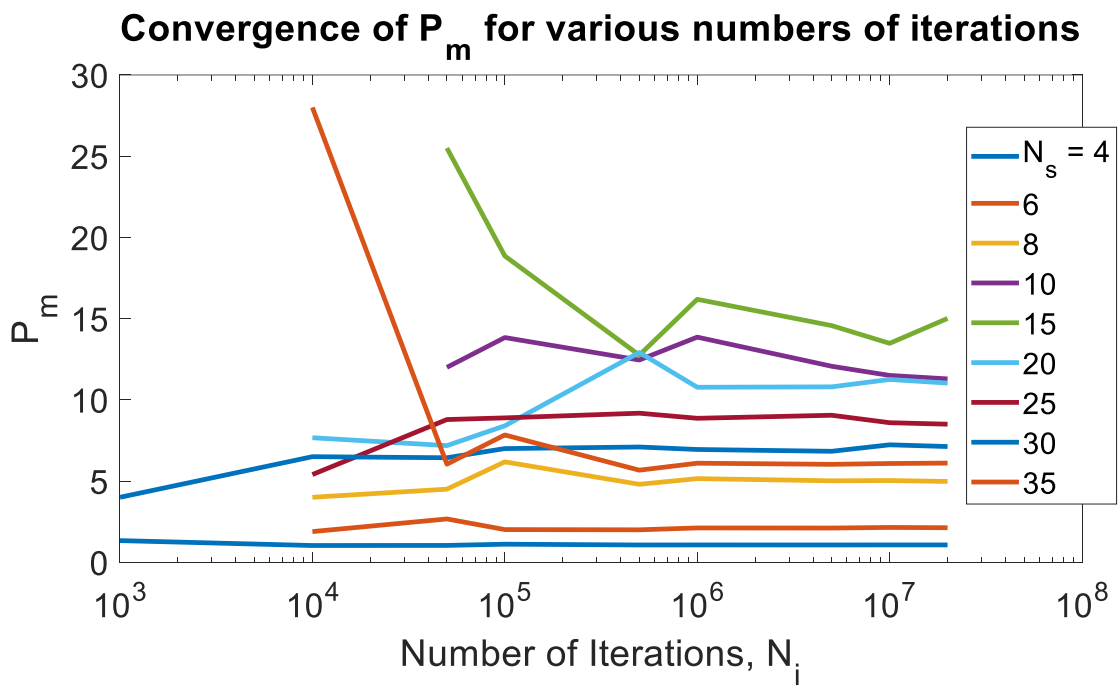

**Figure S5:** Convergence of the metamer reduction metric ($P_m$) as a function of $N_s$ and number of iterations, $N_i$

Thin-film filter design

The filter response design goal (Fig. S3(a)) used in this work is realized by conventional thin-film design methods. A commercial thin-film design software (Optilayer) was used to optimize a two-material thin-film stack to adequately meet the design goal. Tantalum oxide ($Ta_2O_5$) was chosen as the high index (n = 2.15) material and silicon dioxide ($SiO_2$) was chosen as the low index (n = 1.46) material, as they are both easily deposited. The substrate was NBK7, a common optical glass. The final stack was constrained to be less than 75 total layers to keep costs down, and each layer between 10 – 500 nm to prevent stress cracks in thick films. Using these constraints in tandem with the provided filter design goal, the thin-film stack was optimized for incident angles between 0 - 10°. A representative stack design produced by Optilayer for the design of Filter 2 is given in Fig. S5.

The actual design for the device experimentally demonstrated in the main text was slightly modified from that of Fig. S6 by a thin-film foundry (Iridian Spectral Technologies, Ontario, Canada), though they did not share the precise thicknesses of the films with us due to their standard disclosure policy. Nevertheless, the specifics of the design are not critical as long as it implements the desired transmission spectrum (Fig. 2(b)).

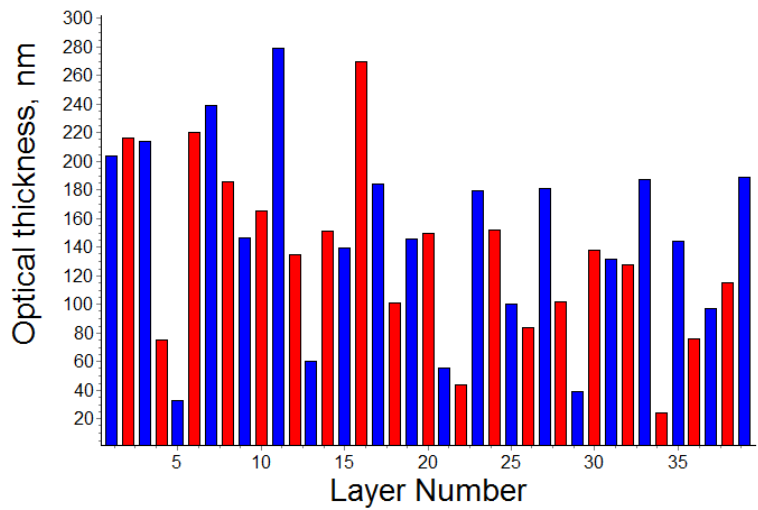

**Figure S6:** Thin-film filter stack design for filter 2 (Fig. S3(a)), using $Ta_2O_5$ (n = 2.15, blue) and $SiO_2$ (n = 1.46, red) dielectric layers.

Comparison of CRT and LCD monitors:

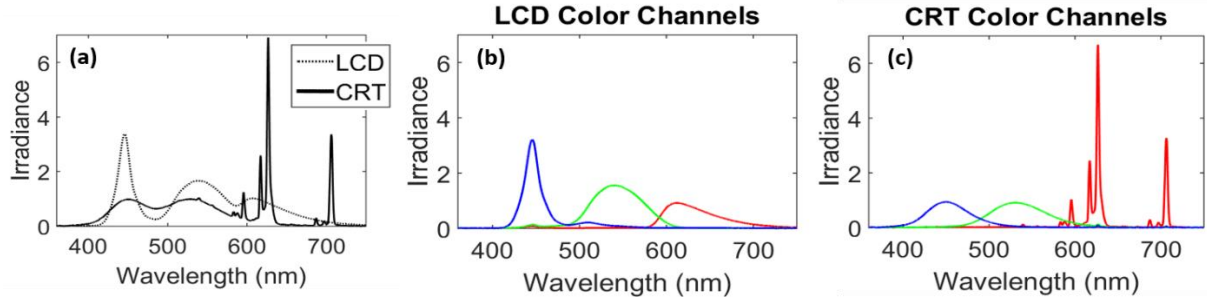

**Figure S7:** (a) Measured emission spectra of the LCD and CRT displays used in this work, displaying a white color (RGB = 255, 255, 255). (b, c) Measured emission spectra of each individual color channel (R, G, B) for the LCD and CRT display, respectively. For blue curves, the displayed color was RGB = (0, 0, 255), red curves RGB = (255, 0, 0), and green curves RGB = (0, 255, 0)

The displays used in this work to generate metameric spectra were a True HD-IPS liquid crystal display (LCD) on a LG G3 smartphone and a conventional cathode ray tube (CRT) monitor (Dell E770P). These displays use very different mechanisms to generate colors, which results in significantly different spectra when displaying the same color (i.e. metamers). The LCD display uses a backlight, typically a white LED, which is transmitted through color filter arrays (red, green and blue color filters) to produce its color response. Therefore, the emitted spectrum is the product of the LCD backlight and color filter transmission response. The CRT monitor uses an electron gun, and relies on a phosphorescent screen to control its spectrum in the visible wavelength range. Because the two display types use significantly different methods to generate colors, the two emission technologies have different spectral features for the individual red, green and blue color channels. The distinct features of the two displays are demonstrated in Figure S7(a), which shows the measured emitted spectrum of white light from each display (RGB = [255, 255, 255]). Figures S7(b, c) show the spectrum of each pure color channel (red, green and blue) for the LCD and CRT display, respectively.

Color accuracy of photographs:

In this work, we used digitally generated color samples from spectroscopic data to demonstrate the splitting of a metameric pair using an LCD and CRT monitor (Fig. 3). This method was used because, due to the difference in spectral response between a camera sensor and the human eye, it is difficult to obtain a precisely color-accurate photograph. This difficulty is shown in Figure S8, which shows the original photograph of the experimental setup in Fig. 3(c), an edited photograph that better approximates the actual color, and a digitally rendered color sample showing the "actual color". The colors are rendered using CIE matching functions, as described above; the [X,Y,Z] values calculated using the matching functions and the measured spectrum can then be converted to the sRGB color space, which is the working color space of most computers. Though the sRGB values will be the same across all display devices, the actual displayed color depends on the calibration of the monitor used. Therefore, the rendered color is Fig. S8(c) only represents the perceived color seen in the experiment when the monitor used to view the image has a perfect color calibration.

It is clear that Fig. S8(a) is significantly different than the generated color sample in Fig. S8(c), even though the camera used in S8(a) and the spectrometer used to acquire the sample that generated the color in S8(c) sampled the same light. Fig. S8(b) is the figure used in the main text (Fig. 3(b)), and was edited to more closely represent the color in S8(c) to prevent confusion. The rendered color samples using spectroscopic data represent the perceived colors of both monitors much more accurately.

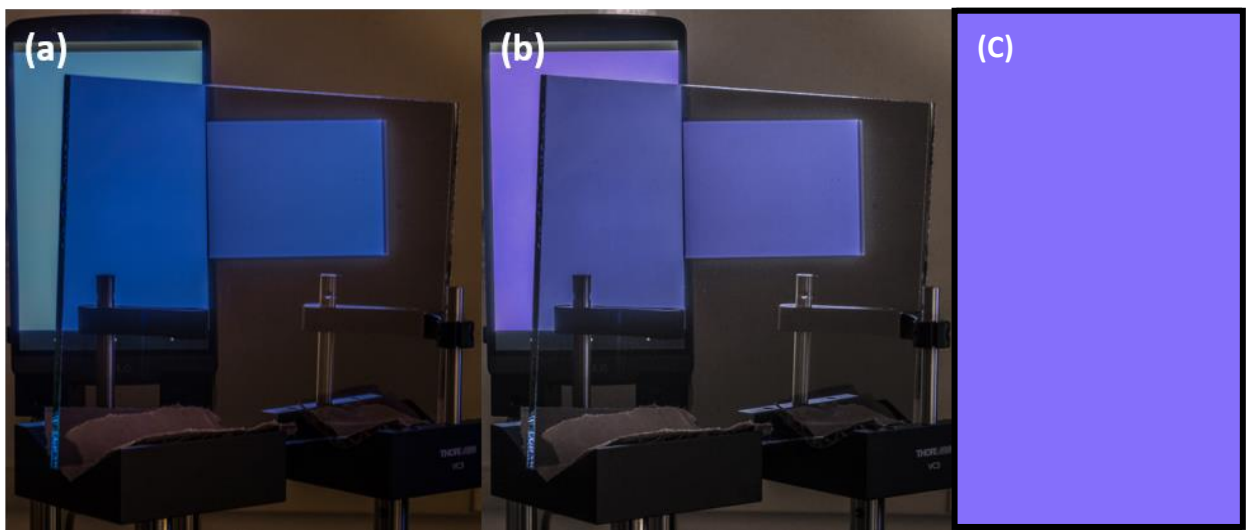

**Figure S8:** (a) Photograph of the setup shown in Fig. 3(c) of the main text, taken using a Sony α7R II camera. (b) The edited photograph that appears in Fig. 3(b) of the main text, which was modified to appear close in color to the rendered colors. This manipulation was performed to prevent confusion in the main text. (c) The actual color displayed during the experiment, rendered using spectra acquired using a grating spectrometer and cosine corrector.

Hyperspectral images

To demonstrate the utility of our wearable passive multispectral device in a more natural setting, we acquired a hyperspectral image of a complex scene, and applied the filters digitally (Fig. S9). The scene included a variety of blue and violet objects, including patches of color made using paints and pastels, plants, and a Morpho butterfly featuring a structural blue color [S5]. The image was obtained using a Middleton Spectral Vision MSV-500 High Sensitivity VNIR hyperspectral camera.

The enlarged images in Fig. S9(d) show the butterfly wing next to six similarly colored samples made using oil pastels, with no filters applied. In Fig. S9€ and Fig. S9(f), filters 2 and 1, respectively, are applied to these enlarged images. The numbers in each panel represent the CIE ΔE color difference between the oil pastel color and the butterfly wing, averaged over a small area to reduce pixel noise. Using filter 2 (Fig. S9(e)), the appearance of the butterfly wing becomes more dissimilar to the pastel samples compared to no filter (*i.e.,* the butterfly "blue" becomes easier to distinguish from the pastel "blue" using the filter). This again demonstrates the effect of partitioning the S cone to provide more spectral information. The improvement is absent for filter 1 (Fig. S9(f)), demonstrating the need for both filters in the design. We note that each filter creates a new set of metamers that may have been distinguishable before; by using two filters, the set of overlapping newly created metamers becomes significantly smaller. Therefore, as long as at least one filter creates an increase in contrast, more spectral information can be communicated to the visual system while decreasing the overall number of possible metamers. Although a slight yellow/green tint is applied to both filtered images, Fig. S9(b) and Fig. S9(c) have similar "white-balance" due to the white-balance condition enforced during the design process.

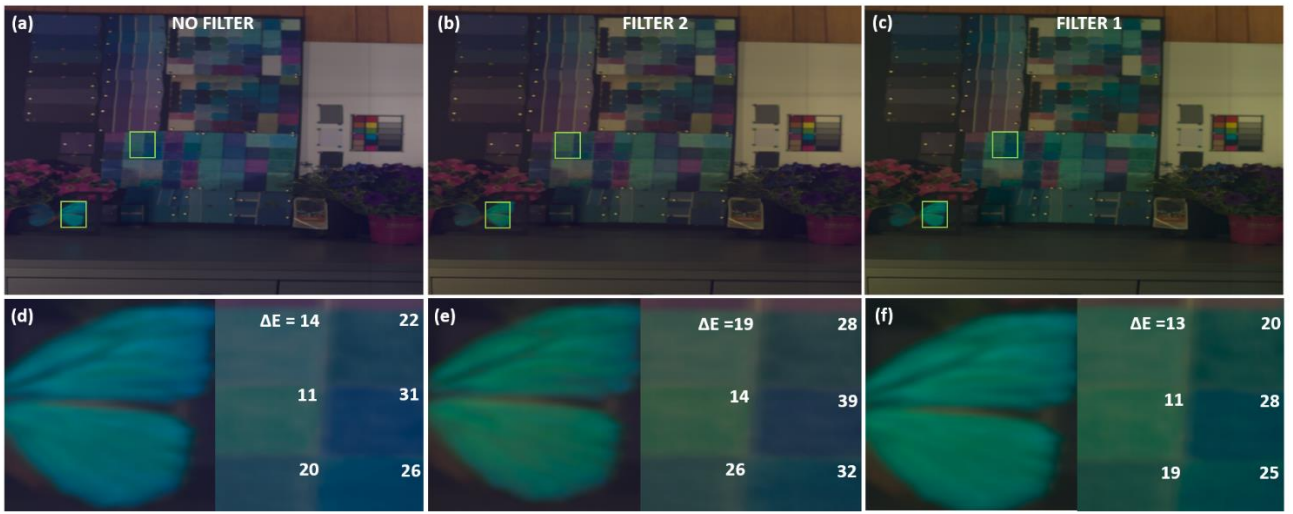

**Fig. S9:** (a) RGB rendering of a hyperspectral image with natural and artificially colored objects, with no filter applied. (b) The render with filter 2 applied, and (c) the render with filter 1 applied. (d) Magnified view of butterfly wing and paint samples (outlined in green in (a)). (e) and (f) are the same samples as in (d), with filters 2 and 1 applied, respectively. The numbers inside each paint sample are the ΔE color difference, rounded to the nearest integer, between the paint sample and the butterfly wing with each respective filter applied.

**Supplementary References**